\documentclass[conference]{IEEEtran}
\IEEEoverridecommandlockouts

\usepackage{float}
\usepackage{soul, color, acronym, multirow, amsmath, balance, amssymb, amsfonts}
\usepackage{cite}
\usepackage{algorithmic}
\usepackage{graphicx}
\usepackage{textcomp}
\usepackage{xcolor}

\def\BibTeX{{\rm B\kern-.05em{\sc i\kern-.025em b}\kern-.08em
    T\kern-.1667em\lower.7ex\hbox{E}\kern-.125emX}}

 \makeatletter
\newcommand{\linebreakand}{%
  \end{@IEEEauthorhalign}
  \hfill\mbox{}\par
  \mbox{}\hfill\begin{@IEEEauthorhalign}
}
\makeatother

\begin{document}

\title{Handover and SINR-Aware Path Optimization in 5G-UAV mmWave Communication using DRL\\
\thanks{Work supported by the European Union under the Italian National Recovery and Resilience Plan (PNRR) PE00000001 - program "RESTART" and by the HORIZON-CL4-2021-SPACE-01 project "5G+ evoluTion to mutioRbitAl multibaNd neTwORks" (TRANTOR) No. 101081983}
}

\author{\IEEEauthorblockN{Achilles Kiwanuka Machumilane}
\IEEEauthorblockA{\textit{Institute of Information Science and Technologies (ISTI)} \\
\textit{CNR}\\
Pisa, Italy \\
achilles.machumilane@isti.cnr.it}
\and
\IEEEauthorblockN{Alberto Gotta}
\IEEEauthorblockA{\textit{Institute of Information Science and Technologies (ISTI)} \\
\textit{CNR}\\
Pisa, Italy \\
alberto.gotta@isti.cnr.it}
\linebreakand 
\IEEEauthorblockN{Pietro Cassar\'a}
\IEEEauthorblockA{\textit{Institute of Information Science and Technologies (ISTI)} \\
\textit{CNR}\\
Pisa, Italy \\
pietro.cassara@isti.cnr.it}
}

\maketitle
\begin{abstract}
Path planning and optimization for unmanned aerial vehicles (UAVs)-assisted next-generation wireless networks is critical for mobility management and ensuring UAV safety and ubiquitous connectivity, especially in dense urban environments with street canyons and tall buildings. Traditional statistical and model-based techniques have been successfully used for path optimization in communication networks. However, when dynamic channel propagation characteristics such as line-of-sight (LOS), interference, handover, and signal-to-interference and noise ratio (SINR) are included in path optimization, statistical and model-based path planning solutions become obsolete since they cannot adapt to the dynamic and time-varying wireless channels, especially in the mmWave bands. In this paper, we propose a novel model-free actor-critic deep reinforcement learning (AC-DRL) framework for path optimization in UAV-assisted 5G mmWave wireless networks, which combines four important aspects of UAV communication: \textit{flight time, handover, connectivity and SINR}. We train an AC-RL agent that enables a UAV connected to a gNB to determine the optimal path to a desired destination in the shortest possible time with minimal gNB handover, while maintaining connectivity and the highest possible SINR. We train our model with data from a powerful ray tracing tool called Wireless InSite, which uses 3D images of the propagation environment and provides data that closely resembles the real propagation environment. The simulation results show that our system has superior performance in tracking high SINR compared to other selected RL algorithms.
\end{abstract}

\section{Introduction}
In recent years, interest in unmanned aerial vehicles (UAVs) has increased significantly, with applications in military operations, transportation, reconnaissance, search and rescue missions, environmental monitoring, photography and telecommunications \cite{khawaja2019survey}. The attractiveness of UAVs lies in their lightweight design, cost efficiency in production and operation and the ability to navigate difficult flight spaces. In non-payload mode, UAVs are used to exchange critical safety information with remote controllers, while in payload communication, UAVs are used to transmit data such as aerial imagery and high-speed video to and from various ground devices \cite{khawaja2019survey,bacco2022air,manlio2021unmanned, bacco2020simulation}. Therefore, ensuring seamless and secure UAV operations as well as reliable and efficient communication between the UAV and its control unit or application server is of utmost importance.
In its Release 15 \cite{3GPP-15UAV}, the 3rd Generation Partnership Project (3GPP) has confirmed the feasibility of supporting UAVs with terrestrial Long-Term Evolution (LTE) and New Radio (NR). UAVs can either act as users in the air accessing the cellular network from the sky, which is referred to as \textit{cellular-connected UAV communication}, or be used as airborne base stations (BS) or relay units, which is referred to as \textit{UAV-assisted cellular communication} \cite{zeng2019accessing}.

Despite their great potential and capabilities, UAVs pose a number of challenges for terrestrial cellular systems, as both LTE and NR were originally developed for terrestrial users only. For example, the high interference resulting from the variations in LOS probability leads to a lower SINR for airborne user equipment (UEs) compared to their terrestrial counterparts  \cite{zeng2019accessing}. Moreover, the high mobility of UAVs often results in more frequent handovers and handover failures, as well as fluctuating wireless backhaul links between UAVs and ground base stations or users, leading to practical difficulties in ultra-reliable low-latency communication (URLLC) scenarios \cite{bacco2014radio}.
Although the NR with multiple-input-multiple-output (MIMO) and beamforming \cite{khawaja2019survey} can provide better performance, further improvements in mobility management are needed as the NR was not originally designed for aerial users. One of the solutions proposed by 3GPP \cite{3GPP-15UAV} is the efficient planning and optimization of the UAV flight path, as this can improve LOS availability, minimize interference and handovers, and ensure reliable connectivity.

Conventional optimization algorithms focus primarily on minimizing flight duration. However, with the stringent quality of service (QoS) requirements envisaged in the upcoming sixth generation (6G) \cite{machumilane2023towards} networks, a UAV must take a route regardless of whether it is the most direct or the shortest route. 3GPP provides that the network can request a terminal to periodically send its path information, which includes a set of location points and associated timestamps that the network can use to predict the sequence of cells that the terminal can visit and the arrival time of the UAV in the network coverage \cite{wu2021comprehensive}. However, such a priori flight planning is only possible if physical features such as buildings and hills are taken into account, as these are static. In such cases, path optimization can be performed using statistical or model-based path optimization methods \cite{zhang2018cellular}. However, if dynamic wireless channel characteristics such as bandwidth, delay, congestion, loss and SINR are to be considered, advanced path optimization methods that can account for the dynamic nature of propagation channels are required. Artificial intelligence (AI)/machine learning (ML) techniques, especially RL, have shown great potential for path optimization in dynamic and time-varying environments as they can adapt to the rapidly changing channel conditions in real time \cite{machumilane2022actor, machumilane2022path}. 

In this work, we propose an AC-DRL-based UAV path optimization scheme in UAV-assisted 5G mmWave communication networks that jointly optimizes \textit{time, handover, and SINR (connectivity)}. We train an RL agent that helps a UAV connected to a gNB to determine the optimal path to a desired destination in the shortest possible time with minimal gNB handover, while maintaining connectivity and the highest possible SINR. We train our AC-DRL model with channel data obtained using a very powerful ray tracing tool called Wireless InSite \cite{WirelessInSite}. This tool takes as input the 3D image of the investigated area and simulates the propagation where the signal interacts with the surrounding geometry, including reflections and diffraction, as it would be the case in the real world. The result is that the obtained channel data is very close to the real data and ensures an accurate and reliable model for path optimization. More details about this tool can be found in Section V.
Our main contributions can be outlined as follows:
\begin{enumerate}
   \item
We present a learning-based framework for UAV trajectory optimization in 5G-UAV communication that jointly optimizes flight time, handovers, and SINR, which have not been combined into a single model in previous studies.
 \item
 Unlike other studies, we train our AC-DRL model with channel data obtained using a ray tracing tool called Wireless InSite with is very powerful in providing channel data that is very close to the actual data ensuring an accurate and efficient path optimization model.
 \item
 We use a model-free AC-DRL algorithm that does not require any prior information about the propagation environment making our model suitable for real-time path optimization in a time-varying UAV propagation environment.
\end{enumerate}

\section{Related Work}
Several research works have been proposed that use RL-based methods for UAV path optimization. In \cite{mohammadi2023analysis}, a Q-learning framework is proposed for optimizing the trajectory of a UAV serving ground users and the gNB central unit (CU) in MIMO Open-RAN, with the goal of maximizing the overall throughput of the network by using the path loss as a reward for the learning agent. The authors in \cite{bayerlein2018trajectory} use Q-learning and model-free RL to optimize the trajectory of an autonomously flying UAV acting as a BS, with the goal of maximizing the overall transmission rate between the UEs on the ground and the UAV. In \cite{gao2021cellular}, a learning-based UAV trajectory optimization is proposed to minimize the energy consumption of the UAV and avoid failures due to energy depletion before the UAV mission is completed. In \cite{liu2019trajectory}, a Q-learning based multi-agent trajectory design and power control algorithm is proposed, in which the UAV power and trajectory are optimized based on user mobility predictions using Twitter data to maximize the instantaneous total transmition rate while satisfying the rate requirements of users. Our proposed framework uses a model-free DRL AC algorithm that does not require any prior information about the propagation environment or the underlying models. It is suitable for real-time path optimization in UAV cell communication with dynamic and time-varying channel characteristics. We jointly optimize the UAV's flight time, handover and SINR. In addition, we trained our algorithm with data obtained from Wireless InSite, a powerful ray tracing tool that generates realistic data, ensuring a more accurate, reliable and efficient model.

\begin{figure}
\centering
\includegraphics[width=\columnwidth]{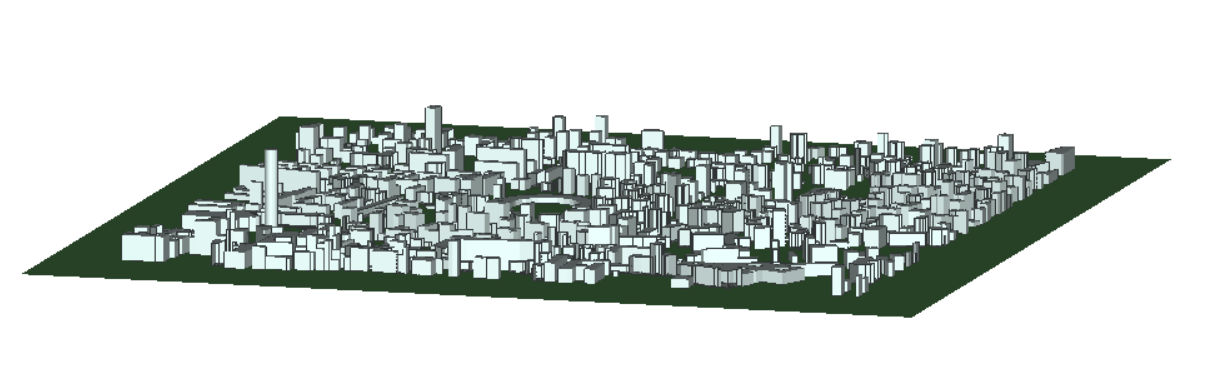}
\caption{Pisa City}
\label{fig:pisa_city}
\end{figure}

\section{System Model}
\subsection{UAV Mobility Model}
We consider a UAV hovering in a grid-like horizontal plane of length $L$ and width $W$ at a constant altitude $H$, as shown in Figure \ref{fig:ref_scenario}. Each grid square is 50m x 50m and the total area is 1km x 1km. The navigation environment consists of a total of $\mathcal{K}$ gNBs fixed at specific locations on the ground with an average Inter-Site Distance (ISD) of 450m which is very close to the 500m ISD provided by 3GPP for urban macro cell (UMa-AV) \cite{3GPP-15UAV}. In this work, the gNB is a transmitter and the UAV is a receiver, but these roles can also be reversed. The goal of this study is to use RL to develop a model that helps the UAV find an optimal path from point $A$ to point $B$ in the shortest possible time while maintaining the highest possible SINR and the least number of handovers. Every 50 m, i.e. in the centre of each square, the UAV makes the decision to move east, west, north or south $(E, W, N, S)$. The UAV position at time $t$ is denoted as $p_t$ = $(x(t), y(t))$. The UAV moves at a constant speed $\nu$ and in time $\tau$, it covers a distance $D_t=\nu \tau$ between the centres of two consecutive grid squares. 

 \subsection{Signal-to-Interference-plus-Noise Ratio (SINR) and Handover }  
 Since the UAV receives signals from different gNBs it experiences interference $I$ from gNBs other than the serving gNB. Therefore, the UAV must perform handovers to gNBs with high SINR. We assume an underlying handover mechanism for UAV-based mobile networks, as provided by 3GPP in \cite{3GPP-18UAV-handover}, which connects the UAV to the gNB with the highest SINR. The signal-to-interference-plus-noise ratio (SINR) is the ratio between the power received by the receiver (UAV) at position $p_t$ as received from the serving gNB and the sum of the power of all interfering transmitters (gNBs) plus the total noise. 
 \begin{equation} \label{eq1}
        SINR(dB)=10 \log_{10}(\Phi_i)-10 \log_{10}(\sigma)-10 \log_{10}(I)
      \end{equation}
where, $ \Phi_i$ is the received power from the serving gNB $i$ while  $I$ is the total interference from other gNBs and $\sigma$, the total noise from all noise sources.

\section{UAV Path Optimization using DLR}
We formulate the UAV path optimization problem as a Markov Decision Process (MDP) and solve it with AC DRL. The UAV connected to a gNB and serving ground users is considered as an agent interacting with the transmission environment by navigating from location $A$ to location $B$ with the goal of maximizing the long-term reward under the following constraints: Reaching the destination via the shortest possible path, maintaining connectivity and the highest SINR with the gNBs, and minimizing the number of handovers from one gNB to another along the path. We define the MDP by the tuple $\{\mathcal{S}, \mathcal{A}, P(s_{t+1}|s_{t},a_t), r_{t}\}$, where $\mathcal{S}$ and $\mathcal{A}$ are the state space and the action space, respectively. $P(s_{t+1}|s_t, a_t)$ is the transition probability from state $s_{t}$ to $s_{t+1}$ as a result of the action $a_t$ performed by the agent at time $t$, and $r_t$ is the agent's reward for the action performed. The state space, the action space and the reward for this problem are described below.
\begin{enumerate}
\item
\noindent \textit{States Space}:
 The state space consists of all possible UAV positions where the UAV has to make decisions about the navigation direction, defined by the (X,Y) coordinates at the position. We refer to the state space as a set of vectors $\mathcal{S}=\{\mathbf{s} \:|\:\mathbf{s} =[ s_{1},\ldots,s_{\kappa}]\}$, where $\kappa$ is the total number of possible UAV positions in the navigation environment and the vector $\textbf{s}_i= [(x_{i}, y_i)]$.

\item
\noindent \textit{Agent's Observation}:
We denote the observation of the agent at time $t$ as a vector $\mathbf{\omega}_t = [(x_t, y_t)]$, where (x,y) are the coordinates at the UAV position at time $t$.

\item
\noindent$Actions$:  We have defined the action space as a vector \textbf{$ \mathcal{A}$}$ = [a_1 \ldots, a_\mathcal{D} ]$ of the four possible directions that the agent can choose,  east (E), west (W), north (N) and south (S).

\item
\noindent$Reward:$  The reward is based on three conditions: the distance from the UAV's position to the destination, the SINR value at the UAV's position as received from the serving gNB, and whether the action performed has triggered a handover to another gNB. The agent's goal is to maximize the long-term reward.

\begin{equation} \label{reward}
 r_{t}=\mu/(d + \rho)
 \end{equation}

where, 
 $d$ is the distance from the UAV position to the destination, $\mu$ is the SINR experienced by the UAV in that position and $\rho=$ is a handover indicator which is 1 if a handover has occurred and 0 if no handover has occurred while the agent transits from $s_{t}$ to $s_{t+1}$
 
 
\end{enumerate}

\subsection*{The AC-DRL Algorithm}
\begin{figure}[!b]
\centering
\centering
\includegraphics[width=0.9\columnwidth]{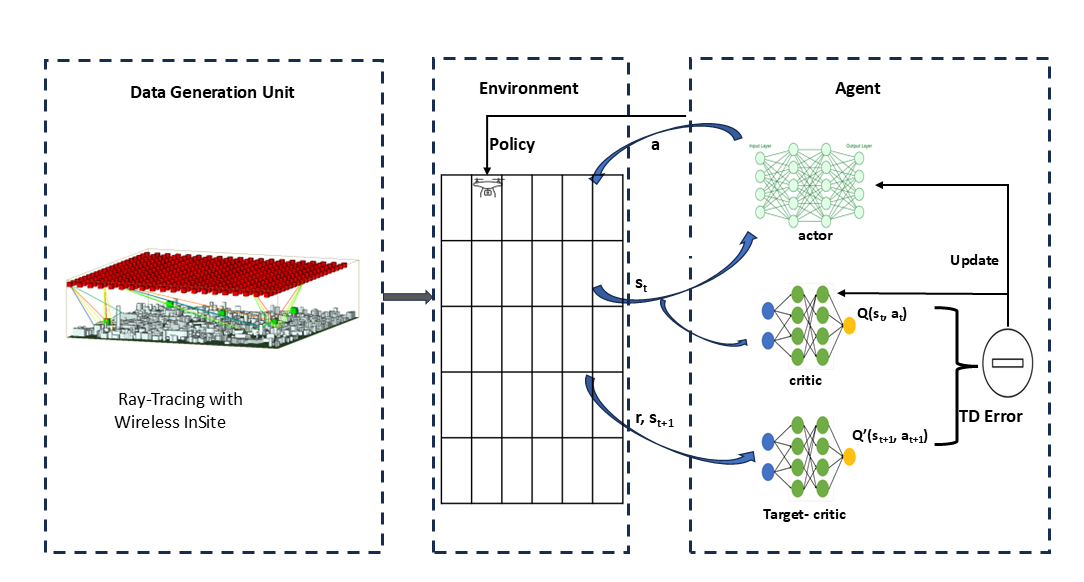}
\caption{System Architecture}
\label{fig:architecture}
\end{figure}

AC is one of the model-free DRL algorithms, which in its basic structure, as shown in Figure \ref{fig:architecture}, consists of the \textit{actor}, which executes the action and uses the policy $\pi(a_t|s_t)$ to achieve the optimization goal, and the \textit{critic}, which calculates the state-action value. The actor updates its policy using the critic's feedback to maximize the overall long-term total rewards. In this work, we use a variant of the AC algorithm called \textit{double AC} that uses two critic networks: the critic that evaluates the immediate state-action value and the \textit{target-critic} that estimates the future rewards that the agent would receive if it were to act with the current policy $\pi(a_t|s_t)$ starting from the current state, also known as Bellman estimation. The target-critic is used to overcome the instability of the critic network caused by frequent updates. The target- critic is updated less frequently than the critic. The policy function $\pi(a_t|s_t)$ can be described with parameters that can be evaluated using various methods including neural networks (NN). In this work, we defined the actor, critic and target-critic networks with parameters $\Theta_a$, $\Theta_c$ and $\Theta_{tc}$, respectively, and evaluated the parameters using fully connected multi-layer perceptron (MLP) NN.
As shown in equation (\ref{eq:actor}), the agent's optimal policy $\pi^*(a_t|s_t)= \displaystyle\arg\max_{a_t}$ can be achieved by the actor maximizing the expected rewards discounted by the factor $\Gamma$.

\begin{equation} \label{eq:actor}
        \pi^*(a_t|s_t)= \displaystyle\arg\max_{a_t} E\Big[{\sum_{t=0}^{\infty}{\Gamma^t r_{t}(s_{t},a_{t})\Big]}}.
    \end{equation}

The critic, calculates the value of the action taken at a given state by estimating the parameters of the state-action function $Q^\pi(s_{t}, a_t)$. The target-critic that estimates the future state-action values is defined as follows: 
\begin{equation} \label{eq:target-critic}
        Q^\pi(s_{t+1},a_{t+1})= E\Big[r_{t}+\Gamma \;\hat{Q}^\pi(s_{t+1},a_{t+1})\Big].
      \end{equation}

The actor is updated by using the temporal difference (TD) error method as given in \cite{grodman2012} as follows:
\begin{equation}
\label{eq:TD} 
\eta  =r_{t}+ \Gamma\hat{Q}^\pi(s_{t+1},a_{t+1})-Q^\pi(s_t,a_t)
\end{equation} 

 The following functions are used to update the actor and critic networks respectively:
 
\begin{equation}
\label{eq:actor-loss}
\delta_{a} = -\lambda\eta \ln\pi(a_t|s_t), 
\end{equation}

\begin{equation}
\label{eq:critic-loss}
\delta_{c} = \beta\eta^2, 
\end{equation}
where $\lambda$ and $\beta$ are actor and critic learning rates respectively. We use the soft-update method to update the target-critic network by copying the weights of the critic network as follows:
\begin{equation} \label{eq:update-target-critic}
    \Theta_{ct}^{new}= \beta\:\Theta_{ct}^{old} + (1-\beta)\Theta_{c}, 
\end{equation}

The AC algorithm was implemented in TensorFlow-2 and  Keras with the ADAM optimizer, 3 hidden layers, 64 neurons per layer, discount factor ($\Gamma$) = 0.96, actor learning rate ($\lambda$) = 0.001 and critic learning rate ($\beta$) = 0.003

 \begin{figure}
\centering
\includegraphics[width=0.9\columnwidth]{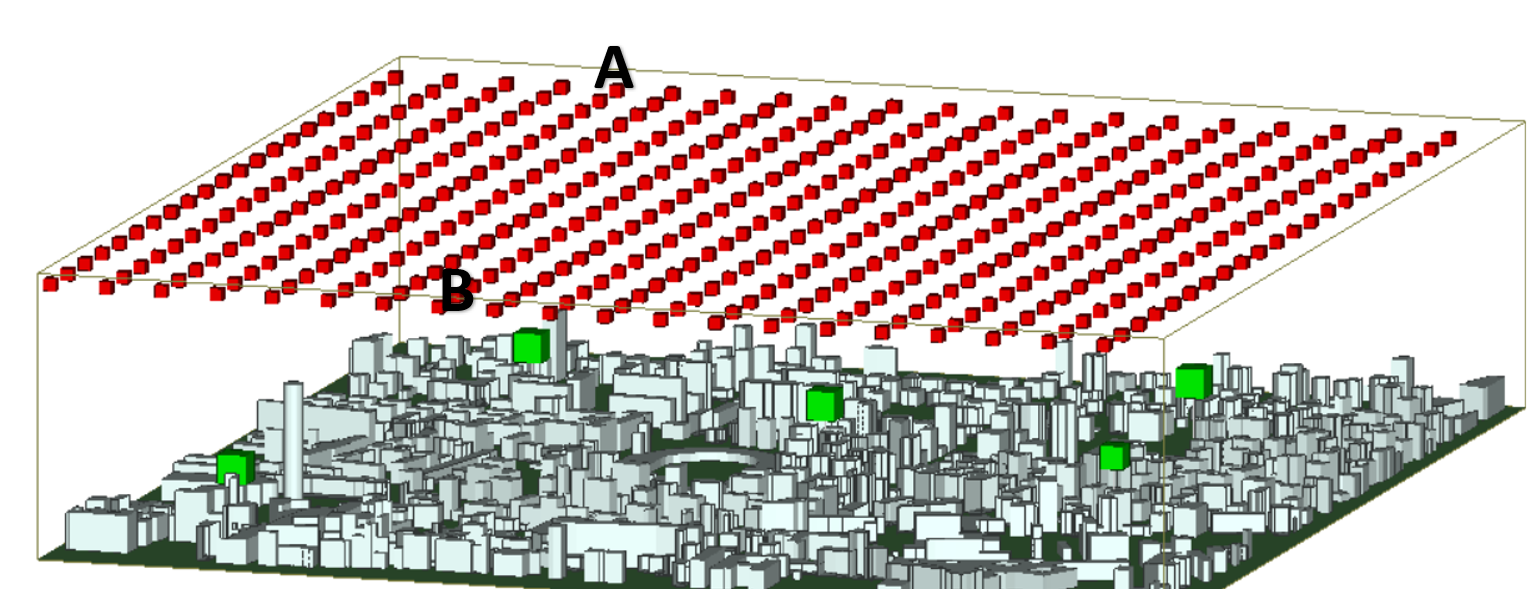}
\caption{Reference Scenario: A grid of the UAV navigation environment on a horizontal plane 150m above Pisa City.}
\label{fig:ref_scenario}
\end{figure}
    
\section{Ray-Tracer Dataset}
\begin{figure}
\centering
\includegraphics[width=0.9\columnwidth]{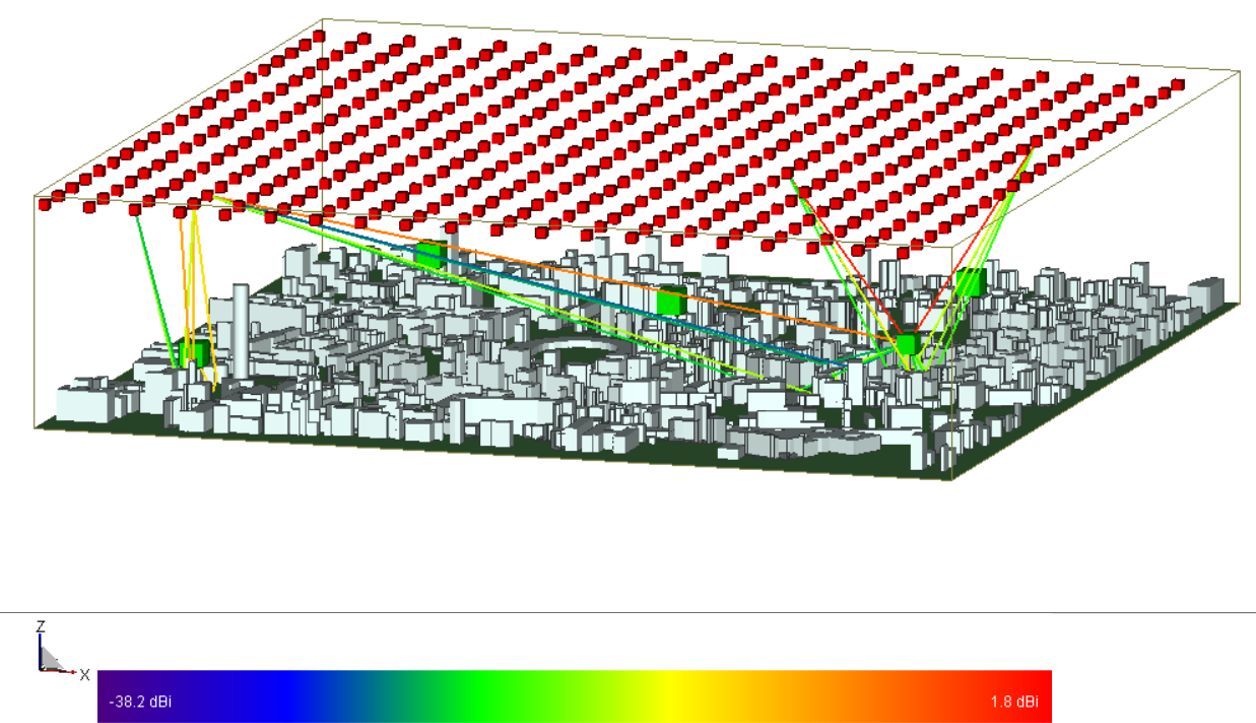}
\caption{A sample of Wireless InSite ray propagation. The green marks show the gNB locations. Notice the reflections and diffractions at surfaces and edges.}
\label{fig:propagarion-rays}
\end{figure}

The scarcity of channel data for training ML models is a significant obstacle in the research on UAVs using ML algorithms, especially in the high frequency domain \cite{8709739}. In this work, we use Wireless InSite \cite{WirelessInSite}, an extremely powerful ray tracing program developed by Remcom. Wireless InSite is a package of Radio Frequency propagation models that provides fast ray-based techniques, empirical models and 3D ray tracing to study site-specific radio wave propagation and wireless communication systems. Thanks to its integrated modeling, simulation and post-processing capabilities, Wireless InSite provides effective and accurate predictions of electromagnetic (EM) radiation propagation and communication channel characteristics in intricate urban, indoor, rural and mixed environments. It provides channel characteristics that are nearly identical to actual conditions, ensuring reliable machine learning models. The rays interact with the geometry of the scene through reflections, diffractions and transmissions as shown in Figure \ref{fig:propagarion-rays}. The electric field, received power and other electromagnetic quantities are calculated and made available for channel analysis. Wireless InSite was also used for channel modeling in \cite{alkhateeb2019deepmimo}. We used this tool to obtain channel parameters for training our model. We used a 1km x 1km 3D map of the city of Pisa in Italy. As can be seen in Figure \ref{fig:pisa_city}, Pisa has buildings with a wide range of heights between 1m and 56m, including the famous Leaning Tower of Pisa, which is 55.86m at the ground on the low side and 56.67m at the high side. As shown in Figure \ref{fig:ref_scenario}, we created a 20 x 20 grid of UAVs with a spacing of 50m making a total of 400 UAVs positions in the propagation environment. We considered the navigation environment in a horizontal plane over the section of Pisa at a constant height of 150m, which is within the flight altitude for UAVs specified by 3GPP \cite{3gpp17}. Five gNBs with a height of 25 m were deployed at specific locations on the ground, as shown in Figure \ref{fig:ref_scenario}. Five radiation paths were transmitted from each gNB to each of the 400 UAVs. At each position, the UAV connects to the gNB that provides the highest SINR at that position while other gNBs are considered as interfering gNBs. From the simulation, we collected the IDs and (X,Y) coordinates of the UAVs and gNBs as well as the highest SINR at each UAV position. Using this data, we trained our AC DRL model to optimize the flight path of the UAV from point A (502361, 945.226) to point B (502661, -4.77445). The simulation results are presented and discussed in the following sections. The ray tracing parameters are shown in Table \ref{ray_tracer}.

\begin{table}
\centering
\caption{Ray Tracing Parameters}
\resizebox{1\columnwidth}{!}{
\label{ray_tracer}
\setlength{\tabcolsep}{3pt}
\begin{tabular}{|p{120pt}|p{60pt}|}
\hline

\textbf{Parameter} &  \textbf{Value}\\
\hline

Number of UAVs and UAV spacing & 400, 50m\\
UAV flying height   & 150m\\
Number of gNBs and gNB height  & 5, 25m\\
Average Inter-Site Distance (ISD) & 450m\\
Building material and Permittivity &Concrete, 7 F/m\\

Building material conductivity & 0.01500 S/m\\

Range of build height & 1m-56m\\
UAV noise figure & 3dB\\ 
Transmit power & 23dB\\
Carrier frequency and bandwidth & 28GHz,  400MHz\\
Velocity of the UAV ($\nu)$ & 10~m/s\\

\hline
\end{tabular}
}
\end{table}

 \begin{figure}[!b]
\centering
\includegraphics[width=\columnwidth]{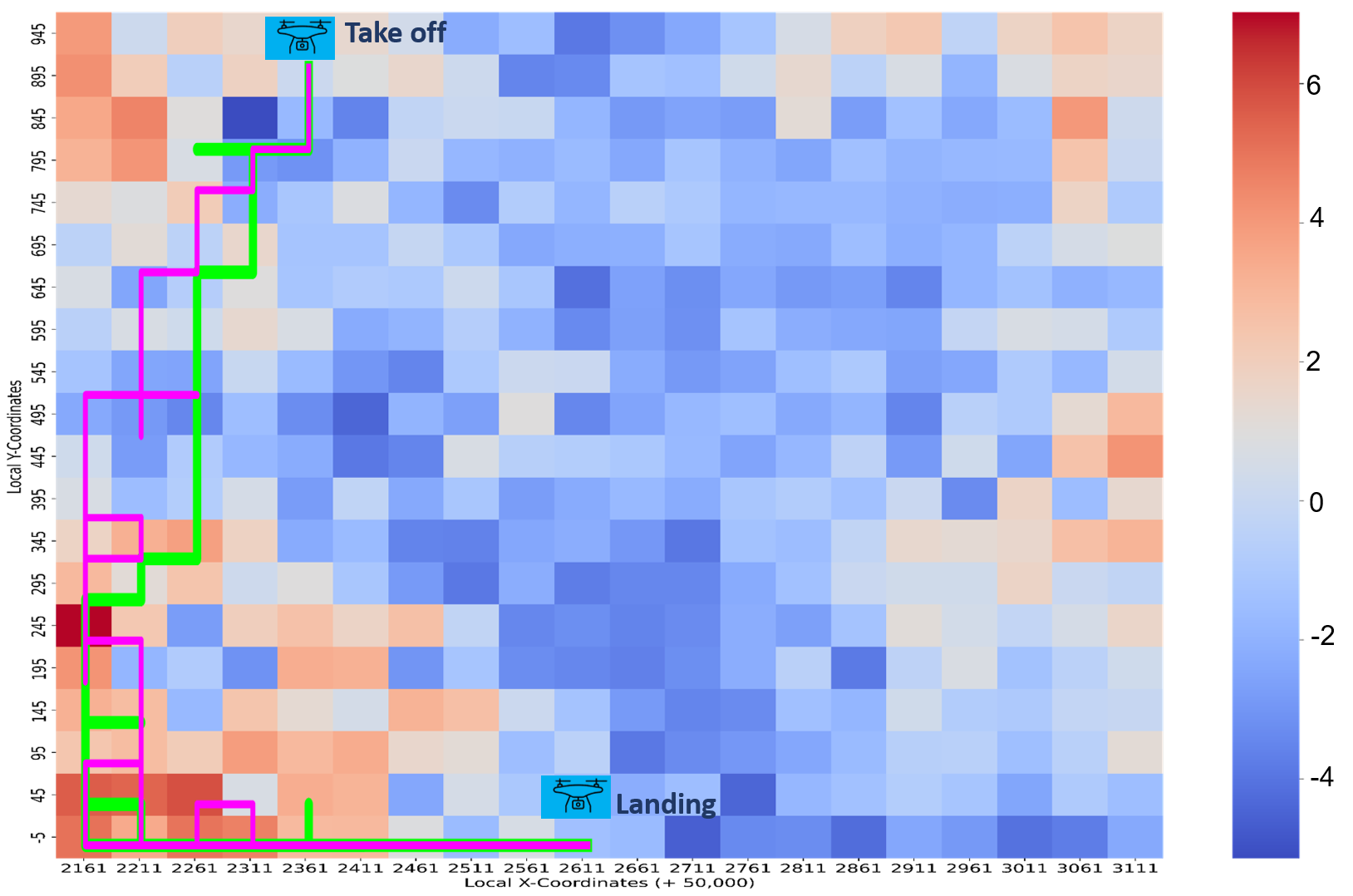}
\caption{UAV-Trajectories using the trained models (green=AC,  magenta=DDQN). The coloured grid shows the maximum SINR at each UAV position in the navigation environment.}
\label{fig:snr-trajectory}
\end{figure}


\section{Performance Evaluation}
In this section, we present and analyze the simulation results to evaluate the performance of our AC-RL algorithm in controlling a UAV in a 5G- UAV network to optimize its trajectory considering three constraints: Using the shortest possible time to move from one point to another while maintaining the highest SINR and minimal handovers. We compare our AC algorithm with another model-free off-policy RL algorithm, the Double Deep Q-Learning Network (DDQN) \cite{van2016deep}. The DDQN is a variant of a DQN, but unlike the DQN, it decouples the selection of the action and the estimation of the action values. Compared to the AC, both the DQN and the DDQN perform better with small action and state spaces, while the AC requires large state and action spaces.

\subsection*{Learning Performance}
\begin{figure}
\centering
\centering
\includegraphics[width=0.95\columnwidth]{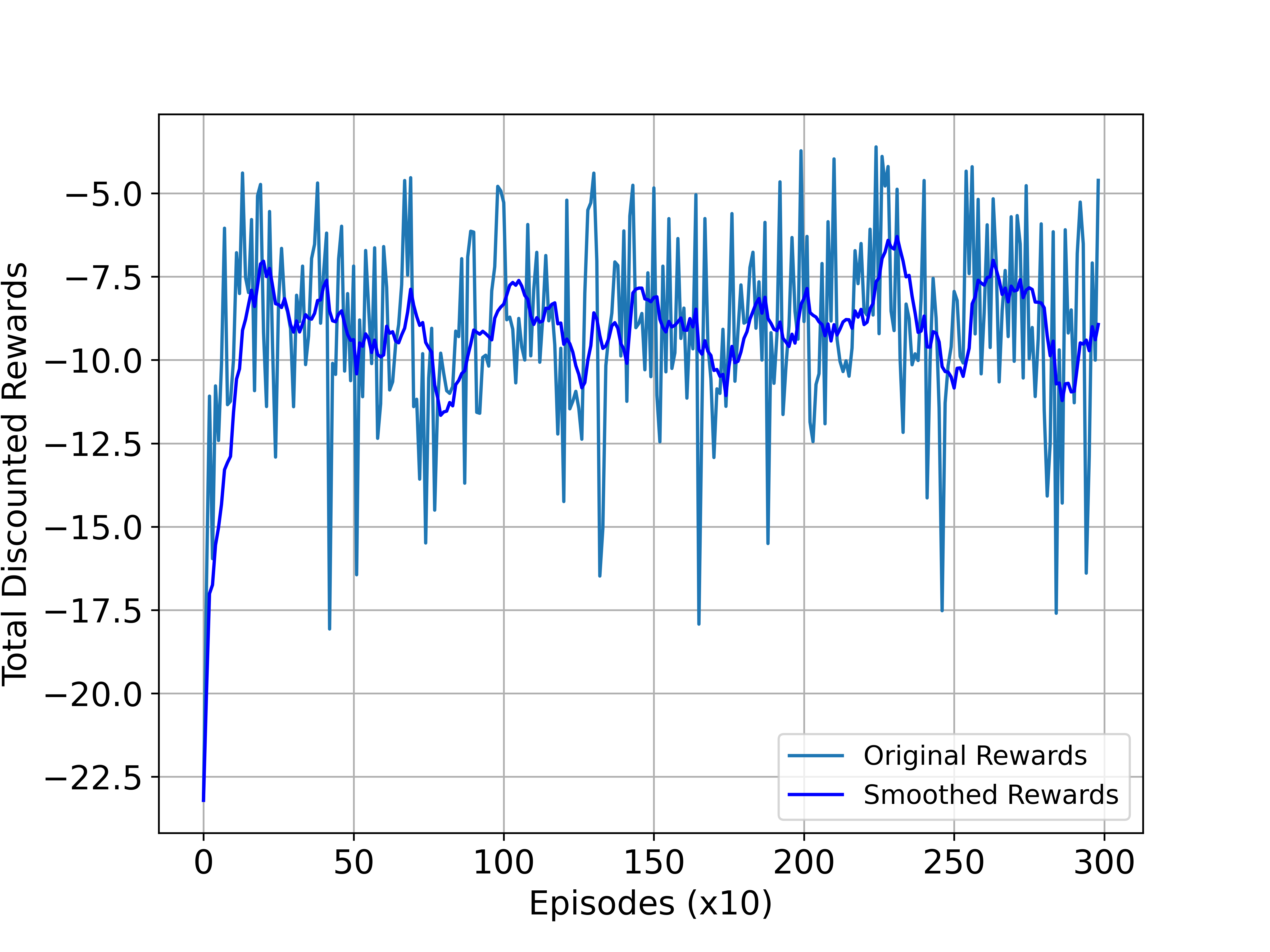}
\caption{Total discounted rewards for the AC-DRL}
\label{fig:rewards}
\end{figure}
Figure \ref{fig:rewards} shows the total discounted rewards achieved by our AC algorithm. The results show that the algorithm performs well and reaches a convergence after a few episodes, allowing the UAV to navigate most of the journey using the optimal policy.

\subsection*{The distance covered and SINR along the UAV Trajectory}
\begin{figure}[!b]
\centering
\includegraphics[width=0.95\columnwidth]{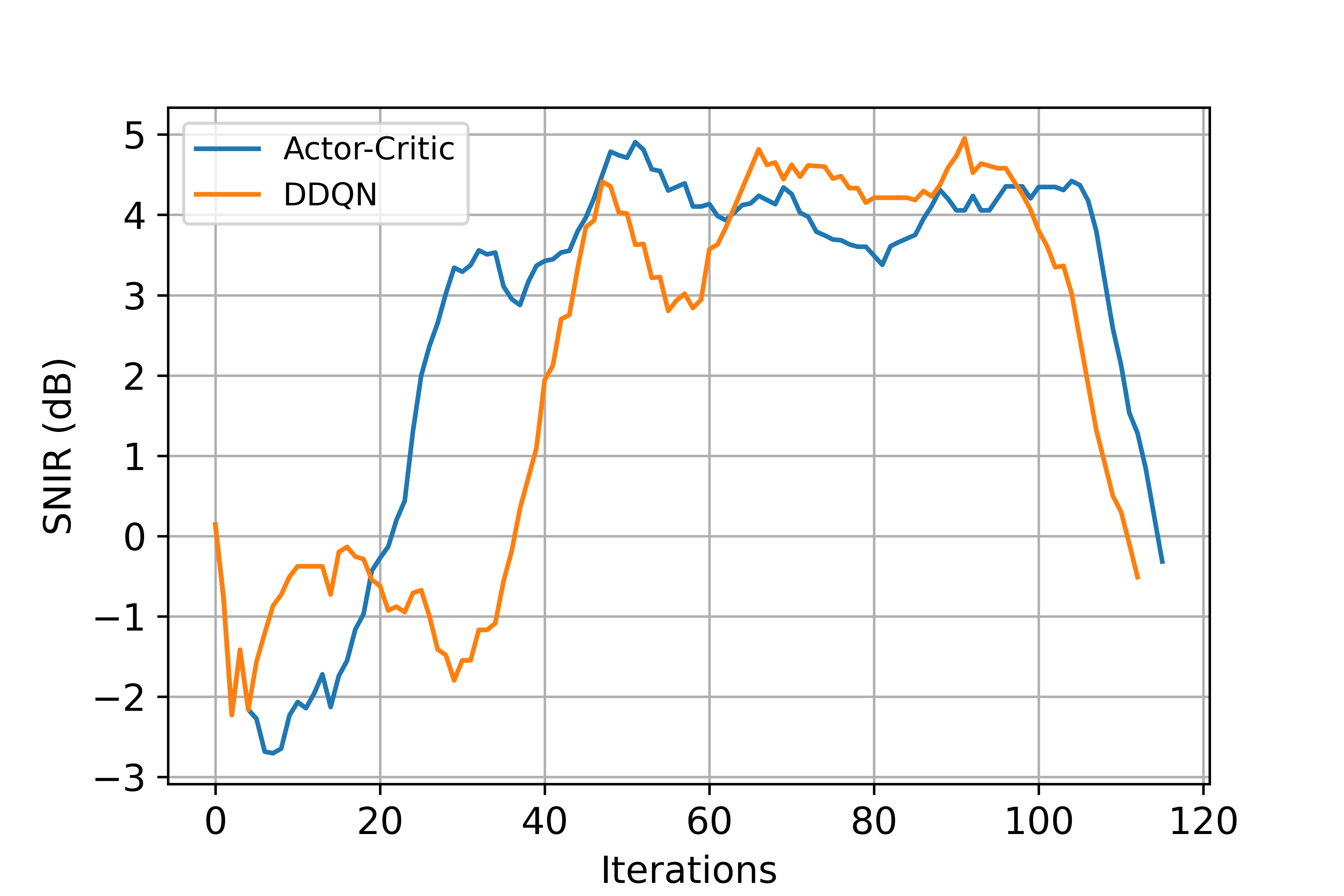}
\caption{SINR values experienced by the UAV along its trajectories using AC and DDQN DRL algorithms.}
\label{fig:snr}
\end{figure}

In Figure \ref{fig:snr-trajectory} we present the optimized trajectories achieved by using the AC and DDQN algorithms. The figure also shows the maximum SINR values at each possible UAV position in the propagation environment grid. As it can be seen from the figure, the central part of the propagation plane receives poor signals, which is probably due to the tall buildings in the city. The two trajectories show that both algorithms have a good and similar performance in following the shortest path with the highest SINR. They have both avoided the central path, which, although is the shortest path, has a low SINR. The right side of the plane was also avoided, even though it has a high SINR, as any path over this side would mean a longer distance to the destination. This shows that the algorithms are able to follow the shortest path with the highest SINR. Table \ref{numerical-results} shows the numerical results for the average SINR, handovers and distance traveled by the UAV using the two algorithms. The results show that the two algorithms performed the same number of handovers, but the AC algorithm achieved a higher SINR compared to DDQN, but at the cost of a 0.1 km longer distance than DDQN.

\begin{figure}
\centering
\includegraphics[width=0.95\columnwidth]{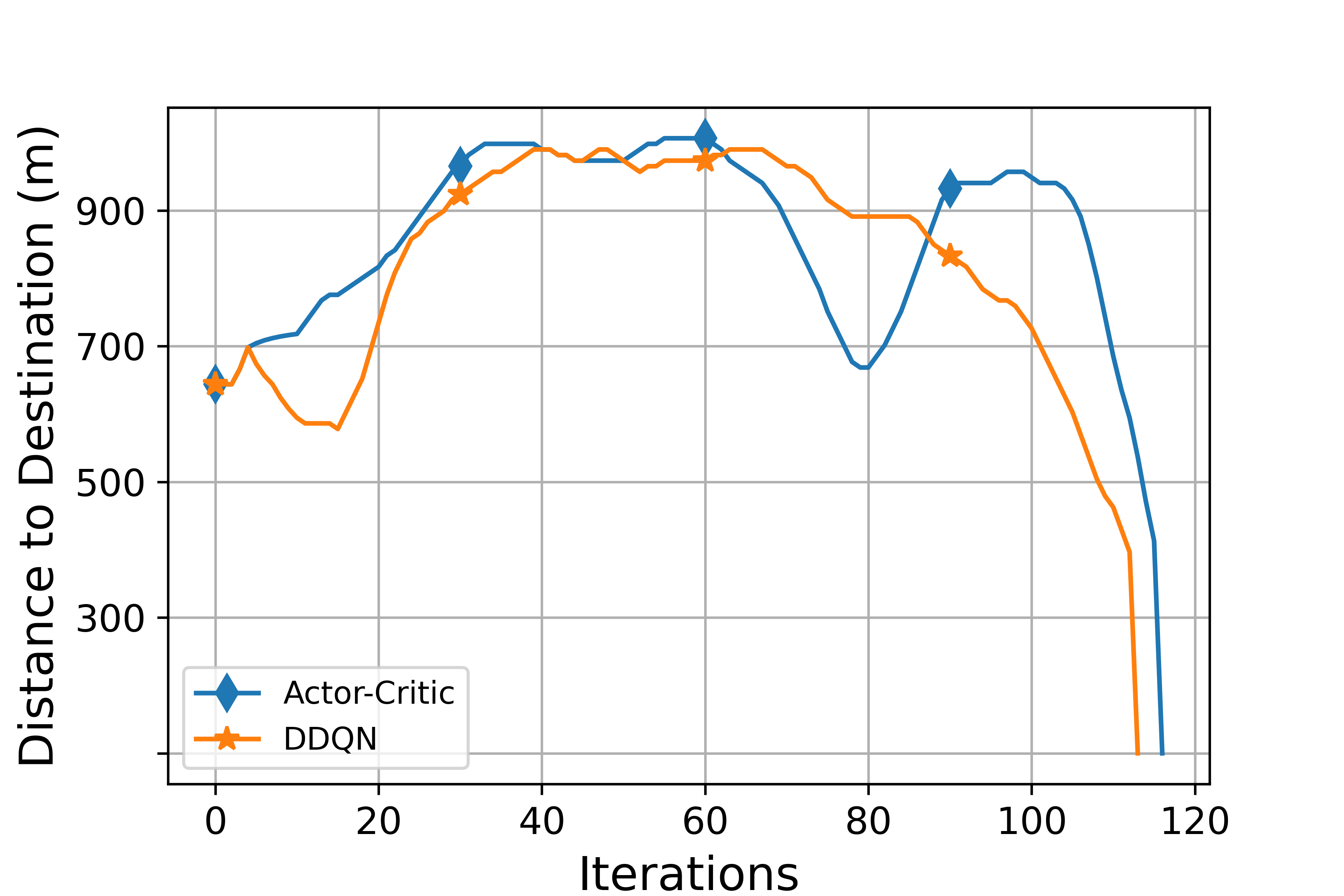}
\caption{The residual distance from each UAV position to the destination  along the trajectories taken using the trained AC and DDQN models.}
\label{fig:distance}
\end{figure}

Figure \ref{fig:distance} shows the distance from each UAV position to the destination. The results show that the distance appears to increase at the beginning, indicating that the UAV is moving away from the destination; i.e., trading off distance in favor of high SINR. Thereafter, the distance towards the destination decreases steadily. Figure \ref{fig:snr} shows the SINR experienced by the UAV along its trajectories. The AC algorithm achieves a slightly higher average SINR of 2.63dB compared to 2.22dB for DDQN, as shown in Table \ref{numerical-results}.

\begin{table}
\centering
\caption{Comparison between AC and DDQN }
\resizebox{1\columnwidth}{!}{
\label{numerical-results}
\setlength{\tabcolsep}{3pt}
\begin{tabular}{|p{60pt}|p{60pt}|p{60pt}|p{60pt}|}
\hline

\textbf{Model}& Average SINR (dB) &  Handovers & Distance Covered (km)\\
\hline

\textbf{AC}   & 2.62618 & 7&5.7\\
\textbf{DDQN}   & 2.2201 & 7&5.6\\
 
\hline
\end{tabular}
}
\end{table}

\section{Conclusion}
\label{sec:conclusion}
In this paper, we have presented an AC-DRL-based framework that jointly optimizes UAV trajectory, SINR, and handover in 5G UAVs mmWave networks. Our AC model shows slightly better performance in terms of achieved SINR compared to DDQN which is expected to perform better due to the small size of the action and state space. To ensure efficiency and robustness, we trained our model with the channel data obtained using a powerful ray tracing tool called Wireless InSite, which uses 3D images of the propagation environment and provides data that is very similar to real propagation environment data. This ensures efficient and reliable model performance in real-world applications. In the future, we plan to further develop our research to consider dual-connectivity scenarios, where the UAV optimizes its route while connecting to two gNBs.
  
\section*{ Acknowledgments}
We  acknowledge Tarun Chawla and Remcom Inc for the Wireless InSite X3D ray tracer used for developing the numerical analysis of this work. 

\bibliographystyle{IEEEtran}
\balance
\bibliography{bibliography/bibliography}
\end{document}